\documentstyle[12pt]{article}
\catcode`\@=11

\font\tenmsa=msam10
\font\sevenmsa=msam7
\font\fivemsa=msam5
\font\tenmsb=msbm10
\font\sevenmsb=msbm7
\font\fivemsb=msbm5
\newfam\msafam
\newfam\msbfam
\textfont\msafam=\tenmsa  \scriptfont\msafam=\sevenmsa
  \scriptscriptfont\msafam=\fivemsa
\textfont\msbfam=\tenmsb  \scriptfont\msbfam=\sevenmsb
  \scriptscriptfont\msbfam=\fivemsb

\def\hexnumber@#1{\ifnum#1<10 \number#1\else
 \ifnum#1=10 A\else\ifnum#1=11 B\else\ifnum#1=12 C\else
 \ifnum#1=13 D\else\ifnum#1=14 E\else\ifnum#1=15 F\fi\fi\fi\fi\fi\fi\fi}

\def\msa@{\hexnumber@\msafam}
\def\msb@{\hexnumber@\msbfam}
\mathchardef\boxdot="2\msa@00
\mathchardef\boxplus="2\msa@01
\mathchardef\boxtimes="2\msa@02
\mathchardef\square="0\msa@03
\mathchardef\blacksquare="0\msa@04
\mathchardef\centerdot="2\msa@05
\mathchardef\lozenge="0\msa@06
\mathchardef\blacklozenge="0\msa@07
\mathchardef\circlearrowright="3\msa@08
\mathchardef\circlearrowleft="3\msa@09
\mathchardef\rightleftharpoons="3\msa@0A
\mathchardef\leftrightharpoons="3\msa@0B
\mathchardef\boxminus="2\msa@0C
\mathchardef\Vdash="3\msa@0D
\mathchardef\Vvdash="3\msa@0E
\mathchardef\vDash="3\msa@0F
\mathchardef\twoheadrightarrow="3\msa@10
\mathchardef\twoheadleftarrow="3\msa@11
\mathchardef\leftleftarrows="3\msa@12
\mathchardef\rightrightarrows="3\msa@13
\mathchardef\upuparrows="3\msa@14
\mathchardef\downdownarrows="3\msa@15
\mathchardef\upharpoonright="3\msa@16

\mathchardef\downharpoonright="3\msa@17
\mathchardef\upharpoonleft="3\msa@18
\mathchardef\downharpoonleft="3\msa@19
\mathchardef\rightarrowtail="3\msa@1A
\mathchardef\leftarrowtail="3\msa@1B
\mathchardef\leftrightarrows="3\msa@1C
\mathchardef\rightleftarrows="3\msa@1D
\mathchardef\Lsh="3\msa@1E
\mathchardef\Rsh="3\msa@1F
\mathchardef\rightsquigarrow="3\msa@20
\mathchardef\leftrightsquigarrow="3\msa@21
\mathchardef\looparrowleft="3\msa@22
\mathchardef\looparrowright="3\msa@23
\mathchardef\circeq="3\msa@24
\mathchardef\succsim="3\msa@25
\mathchardef\gtrsim="3\msa@26
\mathchardef\gtrapprox="3\msa@27
\mathchardef\multimap="3\msa@28
\mathchardef\therefore="3\msa@29
\mathchardef\because="3\msa@2A
\mathchardef\doteqdot="3\msa@2B

\mathchardef\triangleq="3\msa@2C
\mathchardef\precsim="3\msa@2D
\mathchardef\lesssim="3\msa@2E
\mathchardef\lessapprox="3\msa@2F
\mathchardef\eqslantless="3\msa@30
\mathchardef\eqslantgtr="3\msa@31
\mathchardef\curlyeqprec="3\msa@32
\mathchardef\curlyeqsucc="3\msa@33
\mathchardef\preccurlyeq="3\msa@34
\mathchardef\leqq="3\msa@35
\mathchardef\leqslant="3\msa@36
\mathchardef\lessgtr="3\msa@37
\mathchardef\backprime="0\msa@38
\mathchardef\risingdotseq="3\msa@3A
\mathchardef\fallingdotseq="3\msa@3B
\mathchardef\succcurlyeq="3\msa@3C
\mathchardef\geqq="3\msa@3D
\mathchardef\geqslant="3\msa@3E
\mathchardef\gtrless="3\msa@3F
\mathchardef\sqsubset="3\msa@40
\mathchardef\sqsupset="3\msa@41
\mathchardef\trianglerighteq="3\msa@44
\mathchardef\trianglelefteq="3\msa@45
\mathchardef\bigstar="0\msa@46
\mathchardef\between="3\msa@47
\mathchardef\blacktriangledown="0\msa@48
\mathchardef\blacktriangleright="3\msa@49
\mathchardef\blacktriangleleft="3\msa@4A
\def\cobicross{{\triangleright\!\!\!\blacktriangleleft}}

\newcommand{\be}{\begin{equation}}
\newcommand{\ee}{\end{equation}}
\newcommand{\bn}{\begin{eqnarray}}
\newcommand{\en}{\end{eqnarray}}
\newcommand{\bnn}{\begin{eqnarray*}}
\newcommand{\enn}{\end{eqnarray*}}
\newcommand{\ba}{\begin{array}}
\newcommand{\ea}{\end{array}}
\newcommand{\e}{\label}
\newcommand{\r}[1]{(\ref{#1})}

\newcommand{\lra}{\longrightarrow}
\newcommand{\ra}{\rightarrow}
\newcommand{\wt}{\widetilde}
\newcommand{\cU}{{\cal U}}
\newcommand{\cS}{{\cal S}}
\newcommand{\cX}{{\cal X}}
\newcommand{\cH}{{\cal H}}

\newcommand{\cM}{{\cal M}}
\newcommand{\cO}{{\cal O}}
\newcommand{\cP}{{\cal P}}
\newcommand{\ve}{{\varepsilon}}
\newcommand{\os}{{\otimes}}

\newcommand{\rgl}{\rangle}
\newcommand{\lgl}{\langle}
\def\<<{\langle\langle}
\def\>>{\rangle\rangle}

\newcommand{\btr}{\begin{trivlist}}
\newcommand{\etr}{\end{trivlist}}

\newcommand{\poi}{Poincar\'e\ }

\newcommand{\kp}{\kappa}
\newcommand{\one}{{\leavevmode{\rm 1\mkern -5.4mu I}}}
\newcommand{\Ibb}[1]{ {\rm I\ifmmode\mkern
            -3.6mu\else\kern -.2em\fi#1}}
\newcommand{\ibb}[1]{\leavevmode\hbox{\kern.3em\vrule
     height 1.2ex depth -.3ex width .2pt\kern-.3em\rm#1}}
\newcommand{\C}{{\ibb C}}
\newcommand{\R}{{\Ibb R}}
\def\qintd#1#2{\int_{#1}^{#2}\kern-1.5em\raise2.2ex\hbox{${\textstyle
q}$}\ } 
\def\qint{\int\kern-0.9em\raise2.2ex\hbox{${\textstyle
q}$}\ }

\def\larrr{\raise0.3ex\hbox{$\longrightarrow$\kern-1.5em\raise-1.1ex
\hbox{$\scriptstyle{r\rightarrow 1}$}}}
\def\limq{\raise0.3ex\hbox{$\longrightarrow$\kern-1.5em\raise-1.1ex
\hbox{$\scriptstyle{q\rightarrow 1}$}}}

\newcommand{\pok}{{{\cal P}^{(2)}_\kappa}}
\newcommand{\fpok}{{Fun_\kappa({\cal P}^{(2)})}}
\newcommand{\tk}{T^{(2)}_\kappa}
\newcommand{\mik}{{{\cal M}^{(2)}_\kappa}}
\newcommand{\fmik}{{Fun_\kappa({\cal M}^{(2)})}}
\newcommand{\ld}{\lambda}
\newcommand{\Ld}{\Lambda}

\begin{document}
\begin{titlepage}
\vspace{1cm}
\begin{flushright}
NPI MSU 97-7/458\\
hep-th/9701079
\end{flushright}
\vspace{1cm}
\begin{center}
\vspace*{1.0cm}

{\Large {\bf Physical Meaning of Quantum Space-Time Symmetries}}

\vskip 1cm

{\large {\bf A.~Demichev}}\renewcommand{\thefootnote}{*}
\footnote{e-mail: demichev@theory.npi.msu.su}

\vskip 0.5cm

Institute of Nuclear Physics,\\
Moscow State University,\\
119899, Moscow, Russia

\end{center}

\vspace{3 cm}

\begin{abstract}
\normalsize

A definition is given and the physical meaning of quantum transformations of
a non-commutative configuration space (quantum group coactions) is discussed.
It is shown that non-commutative coordinates which are transformed by quantum
groups are the natural generalization of the notion of a tensor operator for
usual groups and that the quantum group coactions induce semigroups of
transformations of states of a system.  Two examples of non-commutative
transformations and the corresponding semigroups are considered.

\end{abstract}
\end{titlepage}

\section{Introduction}

Symmetry concepts are of great importance in studying of physical problems:
for model building, classification and essential simplification of
calculations. As is well known, the mathematical background for treating of
symmetries in classical and quantum physics is provided by the group theory
(or by the theory of supergroups). More recently, new and more general
algebraic objects, called quantum groups \cite{Drinfeld,FaddeevRT} (see also,
e.g. \cite{ChaichianDbk} and refs. therein), have attracted much interest.
The mathematical theory of these new objects arose as an abstraction from
constructions developed in the frame of the inverse scattering method of
solution of quantum integrable models. On the other hand, the theory has
proved to be so rich and powerful that it seems natural to apply it to
different problems far beyond the original area, in particular to
generalizations of such a basic physical notion as space-time symmetries.
This is obviously necessary if space-time is supposed to be described by
non-commutative coordinates.  The construction of relativistic quantum theory
in the frame of "quantum geometry", i.e. geometry with non-commuting and/or
discrete coordinates is an old dream of physicists
\cite{Snyder,Kadyshevsky,Gol'fand} (see also \cite{Blokhintsev} and refs. 
therein), the reason being the hope to obtain in this way quantum theory with
improved ultraviolet properties and new insight on the underlying space-time
geometry. Though these attempts to construct such models were not very
successful because of lacking of corresponding deformed relativistic symmetry
principles, it is generally believed that the picture of space-time as a
manifold should break down at very short distances (of the order of the
Planck length). The uncertainty relations which appear due to gravitational
effects at the Planck scale naturally lead to the conjecture about
non-commutativity of the coordinates \cite{DoplicherFR}.

If one looks for some symmetries in spaces of this type, the first natural
problem is to find (linear) transformations which leave the commutation
relations invariant. However, in general there are no such transformations
with c-number coefficients.  One of the basic ideas of quantum group theory
is to consider transformations of more general type: with matrix elements of
transformations being also non-commutative with each other
\be
x^i\lra x^{\prime\ i}=M^i_{\ j}\os x^j\ ,\quad\quad i,j=1,...,n\ ,\e{coact}
\ee
where both the coordinates $x^i$ and the matrix elements $M^i_{\ j}$ of the
transformation are non-commutative (the tensor product sign stresses that
the coordinates commute with all the matrix elements).

Mathematical theory of these objects (quantum groups and quantum spaces)
has been well developed by now (for a review see, e.g. \cite{ChaichianDbk}).
The physical meaning of \r{coact}, in the case of interpretation of $x^i$
as space-time coordinates, is still to be understood. The straightforward
attempts made in plenty of works to consider transformations induced by
\r{coact} on the level of ``fields'' $\phi(x)$ and to construct
corresponding covariant equations of motion, etc. have not brought much
understanding because the very notion of (quantum) fields depending on
non-commutative coordinates is not well defined. 
An interesting discussion of the problem of a correct definition
and possible consequences of quantum space-time symmetries
is contained in \cite{deAzcarragaKR}. However, our treatment
essentially differs from that in \cite{deAzcarragaKR}. In
particular, we do not assume that parameters responsible for
non-commutativity of coordinates depend on the Planck constant $\hbar$ and
do not consider that non-commutativity of the algebra of functions on a
quantum group by necessity implies their dynamical character and therefore
brings extra degrees of freedom after the quantum group transformation
\r{coact}. 
 
The problem of construction of representations of the operators entering
\r{coact} important for physical analysis of the non-commutative
trasformations as well as the problem of decomposition of tensor products of
these representations were posed and considered on a number of examples in
the work \cite{ChaichianK}.
In the case of internal symmetries the meaning of quantum group
transformations was discussed in \cite{MackS}.

In this paper we aims at the formulation of the notion of quantum groups of
transformations and space-time symmetry and the derivation of transformations
of states of a system induced by \r{coact}. We shall show that the
cotransformations \r{coact} result in appearance of semigroups of
transformations of representation spaces of algebras of functions on quantum
spaces and that the cotransformations \r{coact} can be understood as the
natural generalization of the well known notion of tensor operators.

Another problem which arise in construction of models based on
non-commutative space-time is to connect the low energy description of
particles in frame of a commutative geometry and the description of particles
at high energy (small distances) feeling q-deformation of the space-time.  To
clarify this point, remind how the analogous situation looks in the
superstring theory \cite{GreenSW}. Low energy particles correspond to string
zero modes. If one considers their scattering, specific for string theory
heavy modes come to the play as intermediate states only, essentially
improving ultraviolet behaviour of amplitudes, but giving negligible
contribution to their finite parts. Thus the existence of superstrings does
not contradict to the low energy phenomenology based on ordinary quantum
field theory.  Unfortunately, an analogous consistent picture for a field
theory based on space-time with non-com\-mu\-ta\-ti\-ve coordinates is
absent.

A natural step (at least at the preliminary stage) towards the understanding
of the relation between the low energy phenomenology and physics in
q-deformed space-time is to reduce the number of non-commuting observables of
a system with non-commuting geometry. In \cite{Demichev} the one possible way
to achieve this has been considered. It was proved that making use of the
so-called quantum group twists, one can choose in the special class of quantum
spaces the coordinate frame with commuting coordinates. In such a frame there
are not serious problems with definition of particle states.  On the other
hand, the differential calculus for these spaces still remains deformed and
this implies lattice-like structure of the theory.

In this paper we shall consider another possibility based on quantum geometry
which corresponds to non-commutative coordinates but commutative components
of momenta. In a sense, this way even more natural because asymptotic states
in quantum field theory are defined usually as eigenstates of momentum
operators.

In the next section we shall start the consideration from the simplest
example of Lie algebra $su(2)$ which we interpret as non-commutative
configuration space. Section 3 contains general definitions and constructions
extracted from the discussion of this simple example. In Section 4 we shall
apply the results to the important example of the deformed Minkowski space
which corresponds to the so-called $\kp$-Poincar\'e group
\cite{LukierskiNRT,MajidR,Zaugg} (for simplicity we shall consider
two-dimensional case).

\section{The toy model: p-top} 

Consider the cotangent fibre bundle $T^*SU(2)$ as a phase space of our
``toy'' model system. Of course, usual classical and quantum mechanical
treatment of this phase space corresponds to the well known classical or
quantum top: points of $SU(2)$ parameterize positions of a solid body
(configuration subspace) and points of fibers parameterize (angular) momenta
(see e.g. \cite{Arnold,LibermannM}). We intend to look at this mathematical
object from another point of view. Let us consider the group manifold $SU(2)$
as the momentum submanifold of the phase space and a cotangent fibre of
$T^*SU(2)$ as the coordinate subspace. From a formal point of view and before
a choice of a Hamiltonian, the very exchange of coordinates and momenta does
not bring essentially new features to the theory. But the physical content of
the model becomes quite different: in particular, the momentum subspace
becomes compact. It is clear, that a quantum field theory, based on such a
phase space of a single particle, would not have any ultraviolet divergences.
On the other hand, a free particle moving with a constant momentum (and
commuting components of the momentum, which are nothing but the group
parameters of $SU(2)$) does not feel global topological properties of the
phase space and hence {\it in-} and {\it out-}states seems to exist in this
theory in the usual sense (we leave the problem of relation between Euclidean
and pseudo-Euclidean field theories in this case and, perhaps, some other
problems aside). Remind that the theories of this type were considered long
ago by Snyder, Kadyshevsky and Gol'fand
\cite{Snyder,Kadyshevsky,Gol'fand}. However, these models were formulated
directly at the level of Feynman rules in quantum field theory and suffered
from the absence of general (symmetry) principles for their construction.  As
is well known, the most important, basic and effective principle for a
construction of physical models is the requirement of an appropriate symmetry
(with respect to Lie group transformations, or infinite dimensional gauge
groups, or supergroups, etc.). As coordinates of the configuration space of
the system under consideration are obviously related to non-commutative Lie
algebra $su(2)$, the transformations are expected to have non-commutative
parameters, i.e. to be a kind of quantum group transformations. Keeping this
in mind, let us discuss possible transformations and their meaning in more
details. For shortness we shall call our system ``p-top''.

A point of the phase space
$T^*SU(2)$ is parameterized by $\xi =(g,\omega)$, where $g\in SU(2)$ and
$\omega$ parameterizes points in the space $su^*(2)$ dual to the Lie
algebra $su(2)$. A cotangent bundle over Lie group admits global left (or
right) trivialization \cite{Arnold,LibermannM} 
$$ (\pi,\lambda):\ T^*SU(2)\lra
SU(2)\times su^*(2)\ , $$ 
where $\pi$ is canonical projection $\pi:\
T^*SU(2)\ra SU(2)$, and $\lambda:\ T^*SU(2)\ra su^*(2)$ for the left
trivialization is defined by 
$$
\lambda(\xi)=\hat{L}_{\pi(\xi)^{-1}}(\xi),\qquad \xi\in T^*SU(2)\ , $$
$\hat{L}$ is lifting of left translations on $SU(2)$ to 
$T^*SU(2)$. Note that $T^*_eSU(2)$ ($e$ is unit element of $SU(2)\,$) is
identified with $su^*(2)$. Right trivialization is defined analogously. 

At this stage a non-commutative object, the Lie algebra $su(2)$, comes to
the play (as the simple compact algebra $su(2)$ has the invariant metric
$g_{ij}=\delta_{ij}$, the difference between $su^*(2)$ and $su(2)$ is not
essential). In our model we identify $su^*(2)$ with the
configuration subspace of the phase space $T^*SU(2)$, and $SU(2)$ as the
momentum subspace. Skipping details of the quantization, we just remind
the known result: quantization of such a system corresponds to the
construction of regular representation of the corresponding Lie group, in
our case $SU(2)$ group. The regular representation of $SU(2)$ is realized
in the Hilbert space ${\cal H}=L^2(S^{(3)})$ with the basis formed by the
rotation matrix elements $D^j_{m_1m_2}(\Omega)\,,\ \Omega\in S^{(3)}\,,\
j=0,1/2,1,...\,;\ \ -j\leq m_1,m_2\leq j$ (see, e.g.
\cite{Vilenkin}). 

Thus after the quantization we have three non-commutative ``coordinates''
$J_i\in su(2)\,,$ $i=1,2,3$ of the configuration space
($[J_i\,,J_j]=i\ve_{ijk} J_k$) and the
(commutative) parameters of $SU(2)$ (i.e., coordinates on $S^{(3)}$),
which correspond to momenta.  At first sight, this may cause the problem:
representations of $su(2)$ are labeled by two numbers, $|j,m\rangle$, i.e.
by eigenvalues of the Casimir operator $J^2$ and the projection $J_3$,
while in the momentum subspace we have three commuting operators and,
hence, vectors in the Hilbert space are labeled by three quantum numbers,
$|\alpha,\beta,\gamma\rangle$. Of course, in all cases one has to choose a
maximal set of commuting operators in the whole algebra of observables
${\cal O}_{SU(2)}$, which contains the universal enveloping algebra ${\cal
U}(su(2))$ of the Lie algebra $su(2)$ and the algebra of functions
$Fun(SU(2))$ on $SU(2)$ as the subalgebras, the multiplication of an
element from ${\cal U}(su(2))$ and an element from $Fun(SU(2))$ being
defined by the action of the $su(2)$ vector fields on $Fun(SU(2))$. In
particular, one can use, together with the left invariant vector fields
$J^L_i$, the right invariant vector fields $J_i^R$. The elements of these
sets are commute with each other $[J_i^L,J_j^R]=0,\ \forall i,j=1,2,3$ and
we can take the set of operators $\{(J^L)^2,J_3^L,J_3^R\}$ to define the
basis vectors in ${\cal H}$. Remind that $(J^L)^2=(J^R)^2$ and that
the operators $J_i^R$ can be constructed of $J^L_i$ and elements of
$Fun(SU(2))$ (for the usual top these two sets correspond to the angular
momenta in laboratory frame and fixed-body frame) \cite{BiedenharnL}.  So
the use of $J^R_i$ does not mean that we have expanded the algebra of
observables. The basis vectors of $L^2(S^{(3)})$, i.e. rotation matrix
elements, are the eigenfunctions of the chosen set $$
(J^L)^2D^j_{m_1m_2}=jD^j_{m_1m_2}\,\quad
J^L_3D^j_{m_1m_2}=m_1D^j_{m_1m_2}\,\quad
J^R_3D^j_{m_1m_2}=m_2D^j_{m_1m_2}\,. $$ 
Thus, if a quantum space is used as the deformed configuration space, one has
to construct the maximal set of commuting operators from the complete algebra
of both coordinate and momentum subspaces.  Although in the present example
this statement sounds almost trivially, let us note that in most works
devoted to quantum space-time geometries, coordinate or momentum subspaces
are studied separately.

By the definition, the Hamiltonian $H$ of a free particle must be
independent on coordinates of the configuration space and be a function of
the $SU(2)$ group parameters only: $H=H(\alpha,\beta,\gamma)$. This means
that it does not depend on values of operators $J^L_i,J^R_j$ 
and we may try to define an analog of translational invariance of our
system, i.e. invariance with respect to addition of the non-commutative
coordinates. Fortunately, in our case this is equivalent (from the
mathematical point of view) to the well known addition of angular momenta
of a quantum system: $J'^L_i=J^L_i+S^L_i$, where $S^L_i$ satisfy the
$su(2)$ Lie algebra relations ($[S^L_i,S^L_j]=i\ve_{ijk}S^L_k$) and are
considered as the (non-commuting) parameters of the quantum transformation
(translation) group. The same equality is fulfilled for the
right-invariant generators:  $J'^R_i=J^R_i+S^R_i$, because they related to
$J^L_i$ by the group transformation \cite{BiedenharnL}:
$J_i^R=gJ^L_ig^{-1}\,,\ \ g\in SU(2)$. 

The operators $J'^L_i,\,J'^R_j$ act in the tensor product space 
$\cH_G\os\cH$, where $\cH_G=L^2(S^{(3)})$ is the Hilbert space of
representation of the algebra of functions on our quantum translation
group, i.e. the algebra generated by the operators $S^{L(R)}_i$. In this
specific case $\cH_G$ and $\cH$ are isomorphic. Thus the
mathematically more correct form for the addition of the non-commutative
operators is to introduce the homomorphism
\be
\delta:\ \cU(su(2))\lra \cU(su(2))\os\cU(su(2))\ ,    \e{1aa}
\ee
where $\delta$ (coaction in the notation of the quantum
group theory \cite{Drinfeld,FaddeevRT,ChaichianDbk}) is defined as follows
\be
J'^{L(R)}_i\equiv \delta(J^{L(R)}_i)=
\one\os J^{L(R)}_i+S^{L(R)}_i\os\one\ .              \e{1b}
\ee
One can continue the chain of such cotransformations:
\bnn
J''^{L(R)}\equiv \delta(J'^{L(R)}_i)&=&\one\os\one\os J^{L(R)}_i\\[2mm]
&+&\one\os S^{L(R)}_i\os\one +S^{L(R)}_i\os\one\os\one\ \in\ 
\cU(su(2))\os\cU(su(2))\os\cU(su(2))\ , 
\enn
etc. The invariance of the p-top with respect to this cotransformations
means that properties of the system do not depend on the way of the
realization of the non-commutative coordinates either as elements of the
algebra $\cU(su(2))$, or as the elements $J',\,J'',\,...$ of the
$\cU(su(2))$ subalgebra of the multiple tensor products
$\cU(su(2))\os\cU(su(2))\os\cdots\os\cU(su(2))$. 

The transformed coordinates $J'^{L(R)}_i$ in \r{1b} have the same commutation
relations and, hence, the same representations as the initial coordinates
$J^{L(R)}_i$. This means that the coaction \r{1aa} induces for the p-top the
map \r{2} $\cS\,:\
\cH_{\cU (su(2))}\os\cH_{\cU (su(2))}\lra\cH_{\cU (su(2))}$ which defines the
transformation of states from representation Hilbert space of the
configuration space algebra $\{J_i\}$ under the action of states from the
representation space of the algebra of functions (generated by $S_i$) of the
non-commutative translation group. In this particular case the map $\cS$ is
constructed with help of the well known Clebsh-Gordan coefficients
$C^{j_1j_2j}_{m_1m_2m}$. The eigenstate
$|j,m_1,m_2\rangle$ of the set $\{(J^L)^2,J_3^L,J_3^R\}$ is transformed under
the action of the eigenstate $|J,M_1,M_2\rangle$ of the set
$\{(S^L)^2,S_3^L,S_3^R\}$ (i.e. the vector from representation Hilbert space
$\cH_{\cU(su(2))}$ of the non-commutative translation group) as follows 
\bn
|\psi'\rangle&=&\cS (|J,M_1,M_2\rangle,|j,m_1,m_2\rangle)\nonumber\\[2mm]
&=&\sum_kC^{Jjk}_{M_1,m_1,M_1+m_1}
C^{Jjk}_{M_2,m_2,M_2+m_2}|k,M_1+m_1,M_2+m_2\rangle\ .     \e{3}
\en
Of course, mathematically this is usual formula for addition of two
vectors of the regular representation. However, in the context of our
consideration we interpret the combination $C^{Jjk}_{M_1,m_1,M_1+m_1}
C^{Jjk}_{M_2,m_2,M_2+m_2}$ as the matrix elements of the operator
$$\cS (J,M_1,M_2)|j,m_1,m_2\rangle:=
\cS (|J,M_1,M_2\rangle,|j,m_1,m_2\rangle)$$

of the transformation of the vector $|j,m_1,m_2\rangle$ under the
(non-commutative) translation with the parameter vector $|J,M_1,M_2\rangle$.
As is seen from \r{3}, only commutative subalgebra $S^L_3,S^R_3$ result in
local transformation of the vector $|j,m_1,m_2\rangle$ (the quantum numbers
$m_1,m_2$ are just shifted to $m_1+M_1,m_2+M_2$) while transformations of the
quantum number $j$ are nonlocal and correspond in
\r{3} to the sum  over $k$ (with the well known triangular condition 
$|J-j|\leq k\leq J+j$ 
for nonzero Clebsh-Gordan coefficients) contrary to the case of
usual commutative translations for which one has
$|x'\rangle=T_a|x\rangle=|x+a\rangle$. 

Let us consider the corresponding transformation of a wave function in the
momentum representation. The wave function 
$\langle\alpha,\beta,\gamma|\psi'\rangle$ can be
rewritten as follows
\begin{eqnarray}
\langle \alpha,\beta,\gamma|\psi'\rangle 
& = & \sum_{j,m_1,m_2}
D^j_{m_1m_2}(\alpha,\beta,\gamma)
\langle j,m_1,m_2|J,M_1,M_2;\psi\rangle \nonumber\\ [2mm] 
& = & \sum_{J',n,l,j}D^j_{M_1+n,M_2+l}(\alpha,\beta,\gamma)
C^{JJ'j}_{M_1n(M_1+n)}
C^{JJ'j}_{M_2l(M_2+l)}\langle J'nl|\psi\rangle\nonumber\\[2mm]
& = & D^J_{M_1,M_2}(\alpha,\beta,\gamma)
\langle \alpha,\beta,\gamma|\psi\rangle\ .     \e{4}
\end{eqnarray}
It is clear that the transformation laws \r{3} and \r{4} imply the
corresponding transformation rules for quantum fields in coordinate or
momentum representations. This, in turn, put the restrictions on possible
forms of Lagrangians. 

Thus, the simple example (p-top) of models with 
non-commutative coordinate invariance, considered in this section, shows
that there is natural generalization of the notion of a configuration space
(or space-time) symmetry
to the case of non-commutative ``parameters'' of the transformations.
While transformations of position operators are defined by the coaction of
a quantum group, the dual space of states is transformed according to the
rules of decomposition of tensor product of representations.

\section{General definition of quantum space-time symmetries}

After the preliminary consideration of the classical Lie group and algebra
as an object with non-commutative symmetry, we proceed to the formulation
of general definition of quantum symmetries. 
Let a quantum group $G_q$ coacts on a non-commutative space
$\cX_q$, i.e. there exists the homomorphic map
\be
\delta\,: \ Fun_q(\cX)\lra Fun_q(G)\os Fun_q(\cX)\ ,  \e{1}
\ee
(the algebra $Fun_q(\cX)$ of functions on $\cX_q$ is the configuration
space subalgebra of the algebra of all operators of the given quantum
system). {\it The system is invariant with respect to the quantum group
transformations if all the properties of the system are independent on the
coaction map $\delta$.} In other words, the algebra $Fun_q(\cX)$ can be
realized as the subalgebra of multiple tensor product
$Fun_q(G)\os Fun_q(G)\os...\os Fun_q(\cX)$ and no measurements can
distinguish the description based on the algebras with different numbers
of the factors $Fun_q(G)$.

At first sight, this definition of symmetry transformations may look quite
unusual but, in fact, it is natural generalization of commutative
transformations. Indeed, usual action of a group $G$ of transformations of
a manifold $\cM$ on a function $f\in Fun(\cM)$ is defined by the equality
\cite{Vilenkin} 
\be
T_gf(x):=f(g^{-1}x)\ ,\qquad g\in G,\ x\in\cM\ .       \e{1a}
\ee
The right hand side of this definition can be considered as the function
defined on $G\times\cM$. In other words, the transformations $T$ defines
the map
$$
T\,:\ Fun(\cM)\lra Fun(G)\os Fun(\cM)\ .   $$ 
More customary map $\phi\,:G\os\cM\ra\cM$ is defined for points of the
manifolds, which play the role of the dual set of states for the
commutative algebra of observables (functions) on usual manifolds.

Returning to the transformations with non-commutative parameters, we 
define the map which is dual to the transformations \r{1} of
observables (operators), i.e.
\be
\cS\,: \cH_{G_q}\os\cH_{\cX_q}\lra\cH_{\cX_q}\ ,          \e{2}
\ee
where $\cH_{G_q}$ and $\cH_{\cX_q}$ are the Hilbert spaces of all 
representations of
the algebras $Fun_q(G)$ and $Fun_q(\cX)$. The duality relation $\<<
A|\psi\>>\,:\cO\os\cH_\cO\ra\C$ between an operator $A$ from some
algebra $\cO$ and a vector $\psi$ from the Hilbert space $\cH_\cO$ of the
representations of this algebra, is defined by mean value of $A$ in the
state $\psi\,:\ \<< A|\psi\>>=\langle\psi|A|\psi\rangle$. The
intertwining operator $\cS$ is implicitly defined by the equation
\be
\<<\delta A|\Psi\os\psi\>> =\<< A|\cS (\Psi,\psi )\>>
\ ,                                                      \e{2a}
\ee
where $\Psi$ is arbitrary vector from $\cH_G$, $\psi$ is arbitrary 
vector from $\cH_\cX$ and $\cS (\Psi,\psi )\in\cH_\cX$. In fact, the usual
definition \r{1a} of the action of (classical, commutative) transformation
groups in the space of functions on some homogeneous manifold $\cM$ also
has the general form \r{2a}. Indeed, in this case the duality relation
between the algebra $Fun(\cM)$ and states, i.e. points of $\cM$, is
defined as follows
$$
\<< f|x\>>=f(x)\ ,\qquad f\in Fun(\cM),\ x\in\cM\ .  $$
The same is true for the group manifold:
$$
\<< T|g\>>=T_g\ ,\qquad T\in Fun(G),\ g\in G\ .  $$ 
Thus \r{1a} can be represented in the form
$$
\<<\delta f|g\os x\>> = \<< T\os f|g\os x\>>
=T_gf(x)                
=\<< f|\cS (g,x)\>>=\<< f|g^{-1}x\>>=f(g^{-1}x)\ , $$
where the third equality follows from \r{2a} and in this special case 
$\cS (g,x)=g^{-1}x$.

The essence of the cotransformations (coaction of a quantum group)
\r{coact} is that the primed operators $x'^i\ (i=1,...,n)$ have the same
commutation relations (CR) as the initial ones $x^i\ (i=1,...,n)$. In the
theory of usual transformations the analogous objects are called tensor
operators.

Remind that if $g\ra D(g)$ is a finite dimensional representation of a Lie
group $G$ in a vector space $V$and $g\ra U_g$ is a unitary representation
of $G$ in a Hilbert space $\cH$, the set $\{T^a\},\ a=1,...,dim\,V$ of
operators in $\cH$ with the property
\be
T'^i=D^i_{\ j}(g)T^j=U^{-1}_gT^iU_g \ ,               \e{3.1}
\ee
($D^i_{\ j}(g)$ is the matrix form of the representation $D$) is called
a tensor operator.

The expression for $T'^i$ via the operators $U_g$ clearly shows that the
transformed components of a tensor operator have the same CR as the
initial ones. The second expression for $T'^i$ via matrix elements $D^i_{\
j}$ is similar to the quantum group coaction. However, essential
difference from the case of a coaction of a quantum group \r{coact} is that
due to commutativity of the matrix elements $D^i_{\ j}$ (this is just
numbers), the operators $T'^i$ act in the same Hilbert space as the
initial ones $T^i$.

To give clear physical meaning to space-time coaction we suggest to
present it in the form analogous to \r{3.1}
\be
{\cS}x'^i={\cS}M^i_{\ j}\os x^j=x^i\cS\ ,          \e{3.2}
\ee
where now, of course, $\cS$ is not an operator in the Hilbert space
$\cH_{\cX_q}$ of representations of the algebra of functions $Fun_q(\cX)$ on
the quantum space $\cX_q$. Instead, $\cS$ defines the map \r{2}
with the concrete form of this map being defined by the duality
relation \r{2a}.
Note that our definition is different from the existent definition of
q-tensor operators \cite{BiedenharnT,RittenbergS} and from the definition
of quantum internal symmetry given in \cite{MackS}.

From \r{2} and the discussion in the preceding section it follows that
the matrix elements of the operator $\cS$ in a chosen bases of
$\cH_{G_q}\os\cH_{\cX_q}$ and $\cH_{\cX_q}$ play the role of generalized
Clebsh-Gordan coefficients (GCGC). If the multiple index (set of quantum
numbers) $\{m\}$ defines basis vectors $\psi_{\{m\}}$ of $\cH_{\cX_q}$,
and the set $\{K\}$ defines basis $\Psi_{\{K\}}$ of $\cH_{G_q}$, one can
write 
\be
\psi'=\cS (\Psi_{\{K\}},\psi_{\{m\}})=\sum_{\{l\}} C^{\{K\}}_{\{m\}\{l\}}
\psi_{\{l\}} \ ,                                           \e{3.5}
\ee
where $C^{\{K\}}_{\{m\}\{l\}}$ are the set of GCGC.

This relation can be rewritten also in the slightly different form
\be
\psi'=\cS_{\Psi_{\{K\}}}(\psi_{\{m\}}) =
\sum_{\{l\}} C_{\{m\}\{l\}}(\{K\})
\psi_{\{l\}} \ ,                                           \e{3.6}
\ee
from which one can see that $\cS_{\Psi_{\{K\}}}$ plays the role analogous
to that of $U_g$ in the formula \r{3.1} for usual groups and
$C_{\{m\}\{l\}}(\{K\}) = C^{\{K\}}_{\{m\}\{l\}}$ is analog of the matrix
representation $D^i_{\ j}(g)$, a state $\Psi_{\{K\}}\in \cH_{G_q}$ being
the analog of a point $g$ on a group manifold. 

One can apply analogous consideration to the very quantum group $G_q$
which coacts on itself
\be
M'^i_{\ j}=\Delta M^i_{\ j}=M^i_{\ k}\os M^k_{\ j}\ ,      \e{3.7}
\ee
This leads to the corresponding transformation of vectors in $\cH_{G_q}$
\be
\Psi'=\cS (\Psi_{\{K\}},\Psi_{\{N\}})\equiv
\cS_{\Psi_{\{K\}}}(\Psi_{\{N\}}) = 
\sum_{\{L\}} C^{\{K\}}_{\{N\}\{L\}} \Psi_{\{L\}} \ ,       \e{3.8}
\ee

Two subsequent coactions of the form \r{3.7} induce composition of the
transformations \r{3.8} and general properties of algebra representations
provide its associativity (or, equivalently, this follows from the
coassociativity of Hopf algebras). 
This means that the transformations \r{3.8}
form the {\it semigroup}. The trivial representation 
$\Psi_{\{0\}}\in\cH_{G_q}$ correspond to the identity transformation. 
However, there is no inverse transformation for arbitrary
$\cS_{\Psi_{\{K\}}}$. This means that the transformations \r{3.8},\r{3.5}
do not form a group.

An important problem is the explicit determination of GCGC. One possible
way to do this is the direct generalization of the procedure in the case of
$su(2)$ Lie algebra (see, e.g. \cite{BiedenharnL}). Applying left and
right hand sides of \r{3.2} to the left and right hand sides of \r{3.5}
correspondingly, one obtains the equality (for the convenience and shortness
we use Dirac bracket notations and drop curly brackets indicating that
$K,m,...$ are multiindices)
$$
\cS (M^i_{\ j}|\Psi_K\rgl x^j|\psi_m\rgl)=\sum_lC^K_{ml}x^i|\psi_l\rgl\ ,
$$
which can be rewritten as follows
\bn
%\lefteqn{
\sum_{L,n}\lgl \Psi_L|M^i_{\ j}|\Psi_K\rgl
\lgl \psi_n|x^j|\psi_m\rgl\cS(|\Psi_L\rgl|\psi_n\rgl)
%\nonumber\\ 
& = &
\sum_{L,n,r}\lgl \Psi_L|M^i_{\ j}|\Psi_K\rgl\lgl \psi_n|x^j|\psi_m\rgl
C^L_{nr}|\psi_r\rgl\nonumber\\
&=& \sum_{l,r}C^K_{ml}\lgl \psi_r|x^i|\psi_l\rgl|\psi_r\rgl\ . \e{3.9}
\en
This gives the consistency equations for GCGC
\be
\sum_{L,n}C^L_{nr}\lgl \Psi_L|M^i_{\ j}|\Psi_K\rgl\lgl \psi_n|x^j|\psi_m\rgl
= \sum_{l}C^K_{ml}\lgl \psi_r|x^i|\psi_l\rgl\ .                \e{3.10}
\ee
The equation \r{3.10} must be completed by the normalization conditions
which follows from the normalization of vectors $|K\rgl$ and $|m\rgl$.

Thus, we conclude that a quantum coaction on a quantum space $\cX_q$ induces
{\it the semigroup} of transformations of the representation space
$\cH_{\cX_q}$ of the algebra $Fun(\cX_q)$ of functions on $\cX_q$.

\section{Transformations on the two-dimensional $\kappa$-space}

In this section we shall consider the structure of transformations
of a representation space $\cH_{\cX_q}$ in the case of the so-called
$\kappa$-deformation of the Poincar\'e group and the corresponding
$\kappa$-Minkowski space \cite{LukierskiNRT,MajidR,Zaugg}. For the sake of
simplicity we will consider two-dimensional case.

The advantage of this type of deformation is that momentum components of a
particle remain commutative. As we discussed in the Introduction, this
allows to define asymptotic states of a particle in the same way as in the
undeformed case.

We start from the short description of the two-dimensional
$\kp$-Poincar\'e group $\cP^{(2)}_\kp$ and the corresponding
$\kp$-Minkowski space.

Technical merit of the $\kp$-deformation is that due to the bicrossproduct
construction \cite{MajidR,Zaugg} it can be formulated in terms of usual
Lie groups and algebras (the bicrossproduct is the appropriate
generalization of the notion of the semidirect product of usual groups to the
case of quantum groups). For the details we refer the reader to
\cite{MajidR,Zaugg,Majid-bk} and here we shall mention only the facts
relevant to our further consideration.

The algebra $Fun_\kp(\cP^{(2)})$ can be constructed of two subalgebras,
$Fun(SO(1,1)$ and $Fun_\kp(T^{(2)})$ making use their bicrossproduct. The
algebra $Fun(SO(1,1)$ is generated by the commuting elements 
$M^{\mu}_{\ \nu}\ (\mu,\nu=0,1)$ 
of $2\times 2$ Lorentz matrix $M$ with the natural Hopf structure 
\bn
\Delta(M^\mu_{\ \nu}) &=& M^\mu_{\ \rho}\os M^\rho_{\ \nu}\ ,\qquad
\ve(M)=\one\ ,\nonumber\\[-2mm]
\ \e{4.1}\\[-2mm]
S(M)&=&\eta^{-1}M^\top\eta\ ,\nonumber
\en
(here $\Delta$ is the comultiplication, $\ve$ is the counity and $S$ is
the antipode; $\eta$ is the Minkowski metric tensor: $\eta=diag\{-1,1\}$).
The pseudo-orthogonality of $SO(1,1)$ gives the relations
$
M^\top\eta M=\eta\ ,              $
i.e.
\be
(M^0_{\ 0})^2-(M^0_{\ 1})^2=1\ ,\quad M^1_{\ 1}=M^0_{\ 0}\ , \quad M^0_{\
1}=M^1_{\ 0}\ ,                                \e{4.2}
\ee
so that $Fun(SO(1,1))$ has only one independent generator, e.g. $M^0_{\ 0}$.

The algebra $Fun_\kp(T^{(2)})$ is generated by two independent elements
with the following relations
\bn
[u^0,u^1]&=&i\kp u^1\ ,\qquad\qquad\qquad \ve(u^\mu )=0\ ,  \e{4.3}\\[2mm]
\Delta(u^\mu)&=&u^\mu\os 1+ 1\os u^\mu\ ,\qquad S(u^\mu)=-u^\mu \ .\e{4.4}
\en
It is easy to recognize in this relations the standard Hopf algebra structure
(cf, for example, \cite{ChaichianDbk}) of the undeformed Lie algebra
$igl(1)$ of inhomogeneous transformations of real line (about this Lie
algebra and the corresponding group $IGL(1,\R)$ see, e.g.
\cite{Vilenkin}).  

The complete Hopf algebra $\cP^{(2)}_\kp=ISO_\kp(1,1)$ is constructed as
the bicrossproduct $\cP^{(2)}_\kp=T^{(2)}_\kp\cobicross SO(1,1)$
\cite{MajidR,Zaugg} with help of the structure map
$$\triangleleft\,:\ Fun(SO(1,1))\os Fun_\kp(T^{(2)})\ra Fun(SO(1,1))$$
which reads as
$$
M^0_{\ 0}\triangleleft u^0=i\kp((M^0_{\ 0})^2-1)\ ,\qquad
M^0_{\ 0}\triangleleft u^1=i\kp M^1_{\ 0}(M^0_{\ 0}-1)\ ,
$$
(this means that $Fun(SO(1,1))$ is a right $Fun_\kp(T^{(2)})$-module)
and the coaction $\beta\,:\ Fun_\kp(T^{(2)})\ra Fun(SO(1,1))\os
Fun_\kp(T^{(2)})$
$$
\beta(u^\mu)=M^\mu_{\ \nu}\os u^\nu
$$
(this means that $Fun_\kp(T^{(2)})$ is a left $Fun(SO(1,1))$-comodule). 
The maps $\triangleleft$ and $\beta$ in frame 
of the bicrossproduct construction
give the following defining relations for the complete Hopf algebra
$Fun_\kp(\cP^{(2)})$ (in addition to \r{4.1},\r{4.3})
\be\ba{l}
[u^0,M^0_{\ 0}]=i\kp((M^0_{\ 0})^2-1)\ ,\\[2mm]
[u^1,M^0_{\ 0}]=i\kp M^1_{\ 0}(M^0_{\ 0}-1)\ ,\ea    \e{4.5}
\ee
$$
\Delta (u^\mu)= u^\mu\os 1 + M^\mu_{\ \nu}\os u^\nu\ ,\qquad
S(u^\mu) = -\eta^{\mu\nu}M^\rho_\nu\eta_{\rho\lambda}u^\lambda
$$
(the comultiplication and antipode for $u^\mu$ in the complete $\pok$ are
different from those in $\tk$, cf. \r{4.4}).

It is amusing that the quantum group $\pok=ISO_\kp(1,1)$ is constructed
from the usual Lie group $SO(1,1)$ and the usual Lie {\it algebra} $igl(1)$.
Correspondingly, the dual quantum universal enveloping algebra
$\cU_\kp(iso(1,1))$ is constructed from the Lie algebra $so(1,1)$ and Lie
{\it group} $IGL(1,\R)$ (the defining relations for $\cU_\kp(iso(1,1))$ can
be found in \cite{LukierskiNRT,MajidR,Zaugg}).

The non-commutative coordinates $x^0,x^1$ of the two-dimensional
$\kp$-Minkowski space $\mik$ have the commutation relations similar to those
for the ``parameters'' of translations $u^\mu$ of $\pok$
\be
[x^0,x^1]=i\kp x^1\ ,                  \e{4.6}
\ee
i.e. form the Lie algebra $igl(1)$. From the discussed structure of $\pok$
it is clear that the conjugate components of momenta parameterize the Lie
group $IGL(1,\R)$. Thus the $\kp$-Minkowski space has the same
general structure as the toy model considered in Section 2 and as the
non-commutative space-time considered long ago by Snyder, Kadyshevsky and
Gol'fand \cite{Snyder,Kadyshevsky,Gol'fand}, i.e. with space-time
coordinates being a Lie algebra and the conjugate components of momenta
parameterizing the corresponding Lie group.
This fact considerably helps in constructing representations of $\fmik$
and, moreover, representations of the complete algebra of coordinates on
$\mik$ and components of momentum: they correspond to the regular
representation of the group $IGL(1,\R)$. The latter is constructed in the
space $\cH_\mik =L^2(\R^{(2)}_\vee)$, of quadratically integrable functions
$f(g)=f(b,a),\ g\in IGL(1,\R)$ defined on upper half-plane $\{a>0,\ -\infty
< b<\infty\}$, the domain of variation of the parameters of $IGL(1,\R)$.
Note that if $IGL(1,\R)$ is considered as the group of transformations of
real line, the parameter $a$ corresponds to dilatations (they form
multiplicative subgroup $\R_+$ of positive numbers), and $b$ corresponds to
translations (which form additive subgroup $\R$). Of course, in our case
they have quite another physical meaning of the components of momenta on
$\mik$. 

The (right) regular representation on $\cH_\mik$ is defined by the equality
(cf., e.g. \cite{Vilenkin})
\be
R_{g_0}f(g)=f(gg_0)\ ,                         \e{4.7}
\ee
where the group multiplication has the form
$$
g(b_1,a_1)g(b_2,a_2)=g(b_1+a_1b_2,a_1a_2)\ .
$$
To decompose $\cH_\mik$ to the irreducible components, consider the subspaces
$\cH^\lambda_\mik\,,\ \lambda\in\R$ of functions of the form
$$
f_\lambda(b,a):=\int\,dt\,e^{-it\ld}f(b+t,a)\ ,
$$
with the property
$$
f_\ld(b+b_0,a)=e^{ib_0\ld}f_\ld(b,a)\ .
$$
It is easy to check that the subspaces $\cH^\lambda_\mik$ are invariant
with respect to transformations from $IGL(1,\R)$. It is clear, that for
any $f_\ld\in\cH^\lambda_\mik$ one has 
\be
f_\ld(b,a)=e^{ib\ld}f_\ld(0,a)\equiv e^{ib\ld}\phi(a)\ ,   \e{4.8}
\ee
so that the irreducible representation $R_\ld(g)$ of $IGL(1,\R)$ is
constructed in the space $\cH^\lambda_\mik$ of functions $\phi$ on a 
half-line $0 <\xi <\infty$ and has the form 
$$
R_\ld(g(b,a))\phi(\xi)=e^{\ld b\xi}\phi(a\xi)\ .
$$
(in fact these representations are induced by the one-dimensional
representations of the translation subgroup $T^{(1)}$ of elements $g(b,1)$,
see \cite{Vilenkin}). Moreover, the representations $R_\ld$ and $R_\mu$ are
unitary equivalent if $\ld=p\mu$ for some $p>0$ \cite{Vilenkin}. Thus
there exists only two inequivalent representations of $IGL(1,\R):\ R_+$ and
$R_-$. Arbitrary function $f(b,a)\in\cH_\mik$ is decomposed into the
irreducible components as follows
$$
f(b,a)=\frac{1}{2\pi}\int\,d\ld\,f_\ld(b,a)\ .
$$

The infinitesimal operators corresponding to a representation $R_\ld$ have
the form
\bn
d_R&=&a\partial_a\quad\quad \mbox{(generator of dilatations)}\
,\nonumber\\[-1mm] 
& & \e{4.9}\\[-1mm]
t_R&=&a\partial_b\quad\quad \mbox{(generator of translations)}\ ,\nonumber
\en
or, after the transition \r{4.8} to functions $\phi(\xi)$ of one variable,
\bn
d_R&=&\xi\partial_xi\ ,\nonumber\\[-1mm] 
& & \e{4.10}\\[-1mm]
t_R&=&i\ld\xi\ .\nonumber
\en

As we discussed above, these infinitesimal operators, being the
right-invariant vector fields on $IGL(1,\R)$, play the role of
(non-commutative) coordinates of the configuration subspace. Thus we
identify 
$$
x^0=i\kp d_R\ ,\qquad x^1=i\kp t_R\ .
$$
In the same way as in the toy example of Section 2, to construct complete
set of commuting operators, whose eigenvalues label vectors in $\cH_\mik$
one has to take into the consideration the left-invariant vector fields
$$\ba{lclcl}
d_L&=&a\partial_a+b\partial_b&\quad\quad& \mbox{(left generator of
dilatations)}\ ,\\[2mm] 
t_L&=&\partial_b&\quad\quad& \mbox{(left generator of translations)}\ .
\ea $$
One can immediately see, that the quantum number $\ld$, which distinguish
different subspaces $\cH^\ld_\mik$ is just the eigenvalue of $t_L$.
Diagonalizing one of the right-invariant operators, e.g. $t_R$, so that
$$
t_R\phi_{\ld, k}(\xi)=ik\ld\phi_{\ld, k}(\xi)\ ,
$$
we come to the basis of $\cH_\mik$, labelled by two numbers $\ld, k\in\R$.
From the point of view of quantum mechanics on $IGL(1,\R)$ the functions
$\phi_{\ld ,k}(\xi)$ are eigenfunctions of the coordinate operators and the
existence of the two quantum numbers (``values of coordinates'') corresponds
to the two commuting components $a$ and $b$ of momenta. The scalar product on
$\cH_\mik^\ld$ reads as
\be
(\phi,\psi)=
\int_0^\infty \overline{\phi(\xi)}\psi(\xi)\frac{d\,\xi}{\xi}\ .\e{4.11}
\ee
The right invariance of the measure in \r{4.11} provides unitarity of the
representations $R_\ld$. For the sake of convenience we shall use another
variable $y\,:\ \xi=e^y,\ y\in\R$. The operators of coordinates now take the
form 
\be
x^0=i\kp\partial_y\ ,\qquad x^1=-\kp\ld e^y\ ,          \e{4..12}
\ee
and the scalar product is the usual one on the whole real line
$$
(\phi,\psi)=\int_{-\infty}^\infty \overline\phi(y)\psi(y)\,dy\ .
$$
The next step is the construction of the representations of $\fpok$. The
CR \r{4.3} for translation ``parameters'' $u^0,u^1$ is the same as that for
the quantum coordinates $x^0,x^1$ (see \r{4.6}). The CR \r{4.5} for the
additional operators $M^0_{\ 0},\ M^0_{\ 1}$ (with the relation \r{4.2})
look rather complicated because of the nonlinear right hand sides. To
simplify them and to resolve the constraint \r{4.2} let us introduce the
operator $v$ such that
$$
M^0_{\ 0}=M^1_{\ 1}=\frac{v^2+1}{v^2-1}\ ,\qquad 
M^0_{\ 1}=M^1_{\ 0}=\pm\frac{2v}{v^2-1}\ .
$$
One can check that the operator $v$ has much simpler commutation relations
\bn
[v,u^0]&=&i\kp v\ ,\nonumber\\[-1mm]
 & & \e{4.13}\\[-1mm]
[v,u^1]&=&i\kp\ .\nonumber
\en
Moreover, the operator 
\be
\Ld=\frac12 (u^1v+vu^1)-u^0=u^1v-u^0+i\kp\ ,      \e{4.14}
\ee
commutes with all the operators of the algebra, i.e. this is the central
operator of the algebra \r{4.3},\r{4.13}, and, hence, its eigenvalues defines
different irreducible representations. From \r{4.14} it follows that the
operator $v$ has the general form
\be
v = (u^1)^{-1}(u^0+\wt{\Ld})\ ,\qquad \wt{\Ld}\equiv\Ld-i\kp\,;
\quad \Ld,\kp\in\R                                \e{4.15}
\ee
(we use the same character for the central operator $\Ld$ and for its
eigenvalues). Note that the operator $u^1\sim e^y$ is 
obviously invertible one.

Thus the representations of $\fpok$ are realized in the same Hilbert 
space $\cH=L^2(\R)$ as the representations of $\fmik$ with the only
addition that the representations of the operators from $\fpok$ carry one
more quantum number $\Ld$.

The transformations \r{3.5} in the case of $\pok$ and $\mik$ take the form
$$
\cS(|\Psi_{\Ld_1,\ld_1,y_1}\rgl |\psi_{\ld_2,y_2}\rgl)=
\int\,d\ld_3\,dy_3\,C(\Ld_1,\ld_1,y_1;\ld_2,y_2;\ld_3,y_3)
|\psi_{\ld_3,y_3}\rgl\ ,
$$
where $C(\Ld_1,\ld_1,y_1;\ld_2,y_2;\ld_3,y_3)$ is GCGC for the algebras
$\fpok$ and $\fmik$. Actually, this is the function of the continuous
variables; that is why we write the parameters of the representations 
$\Ld_1,\ld_1,y_1;\ld_2,y_2;\ld_3,y_3$ as arguments of the function and not as
indices (cf. \r{3.5}). Using the bases of eigenfunctions of $u^1$ in
$\cH_\pok$ and $x^1$ in $\cH_\mik$ (i.e. $\delta$-functions) and the general
condition \r{3.10} one obtains two equations for
$C(\Ld_1,\ld_1,y_1;\ld_2,y_2;\ld_3,y_3)$ 
$$
\ba{l}\displaystyle{
\Bigl[\partial_{y_2}(1+2J^{(1)}_+J^{(1)}_-)-
i\ld_2e^{y_2}(J^{(1)}_+ +J^{(1)}_-)+\partial_{y_1}-\partial_{y_3}\Bigl]
C(\Ld_1,\ld_1,y_1;\ld_2,y_2;\ld_3,y_3)=0\ ,}\\[3mm]
\displaystyle{
\Bigl[\partial_{y_2}(J^{(3)}_+ +J^{(3)}_-)-
i\ld_2e^{y_2}(1+2J^{(1)}_+J^{(1)}_-)-i\ld_1e^{y_1}+i\ld_3e^{y_3} \Bigr]
C(\Ld_1,\ld_1,y_1;\ld_2,y_2;\ld_3,y_3)=0\ ,}
\ea
$$
where
$$
J^{(i)}_\pm=\pm\Bigl[1\pm\kp^{-1}\ld_i^{-1}e^{-y_i}(i\kp\partial_{y_i}
- \wt{\Ld}_i)\Bigr]^{-1}\ .
$$
These equations are rather complicated and the explicit 
form of GCGC for $\fpok$ and $\fmik$ will be considered elsewhere.

\section{Conclusion}

We have shown that quantum group cotransformations \r{coact} of quantum
configuration space (or space-time) coordinates can be represented as the
natural generalization (cf. \r{3.2},\r{2}) of transformations of tensor
operators in the usual group theory. This implies, in turn, that the
cotransformations induce the semigroup of transformations \r{3.5} in the
Hilbert space of representations of the algebra of functions on a quantum
space.  These transformations are defined by the generalized Clebsh-Gordan
coefficients, describing decomposition of tensor products of representations
of algebras of functions on quantum spaces and representations of the
corresponding quantum group.  We have considered the interesting example of
two-dimensional $\kp$-Minkowski space and $\kp$-\poi group. The explicit form
of the generalized Clebsh-Gordan coefficients and, hence, of transformations
of Hilbert space of states of a particle in the $\kp$-deformed space-time are
defined by the complicated differential equations. 

The interesting feature of the $\kp$-Minkowski space is that it has the
general structure of the models with non-commutative space-time suggested
long ago \cite{Snyder,Kadyshevsky,Gol'fand}, but with very essential
advantage of having well defined quantum group of (co)transformations, namely
the $\kp$-\poi group. However, we have to note that $\mik$ is based on the
noncompact Lie algebra and the conjugate momenta parameterize the non-compact
Lie group. Thus there is no evidence that $\mik$ results in ultraviolet
finite quantum field theory (because of momentum integration over infinite
volume).  On the other hand, it is known that the realistic compact
space-time can be constructed via conformal compactification (see, e.g.
\cite{VeblenPS,Budinich}). It would be very interesting to find
non-commutative symmetries for such models. This would probably open the way
for construction of ultraviolet regularized theories with adequate space-time
symmetries.
 
\vskip 10 mm
\centerline{{\bf Acknowledgements}}
I would like to thank M.~Chaichian and P.~P.~Kulish for valuable discussions.
This work was partially supported by the INTAS-93-1630-EXT and
RFBR-96-02-16413-a grants.

\end{document}